# Surface Composition of Pluto's Kiladze Area and Relationship to Cryovolcanism

**A. Emran** [1, 2], **C. M. Dalle Ore** [3], **D. P. Cruikshank** [4], **J. C. Cook** [5]

[2] Space and Planetary Sciences, University of Arkansas, Fayetteville, AR 72701, USA.
 al.emran@jpl.nasa.gov
[3] Carl Sagan Center, SETI Institute, Mountain View, CA 94043, USA.
[4] Department of Physics, University of Central Florida, Orlando, FL 32816, USA.
[5] Pinhead Institute, Telluride, CO 81435, USA.

**Abstract**

A link between exposures of water ($H_2O$) ice with traces of an ammoniated compound (e.g., a salt) and the probable effusion of a water-rich cryolava onto the surface of Pluto has been established in previous investigations (Dalle Ore et al. 2019). Here we present the results from the application of a machine learning technique and a radiative transfer model to a water-ice-rich exposure in Kiladze area and surroundings on Pluto. We demonstrate the presence of an ammoniated material suggestive of an undetermined but relatively recent emplacement event. Kiladze lies in a region of Pluto's surface that is structurally distinct from that of the areas where similar evidence points to cryovolcanic activity at some undetermined time in the planet's history. Although the Kiladze depression superficially resembles an impact crater, a close inspection of higher-resolution images indicates that the feature lacks the typical morphology of a crater. Here we suggest that a cryolava water carrying an ammoniated component may have come onto the surface at the Kiladze area via one or more volcanic collapses, as in a resurgent volcanic caldera complex. Large regions east of Kiladze also exhibit the presence of $H_2O$ ice and have graben-like structures suggestive of cryovolcanic activity, but with existing data are not amenable to the detailed search that might reveal an ammoniated component.

## 1. Introduction

The discovery of exposures of water ($H_2O$) ice bearing an ammoniated component on Pluto (Dalle Ore et al. 2019) was the compositional evidence that led to the hypothesis that cryovolcanic activity has shaped the topography in some regions of the planet's surface (Cruikshank et al. 2019). Cryovolcanism, along with the theory of reorientation of Sputnik Planitia (e.g., Nimmo et al. 2016) implies the existence of a subsurface ocean on Pluto and has made this body one of the solar system's ocean worlds. These bodies that have reservoirs of liquid $H_2O$ under their icy crusts in the past and possibly in the present epoch, are prime candidates for presenting the fundamental conditions for fostering life and therefore are among the contenders for future *in situ* space missions (e.g., Cruikshank et al. 2019b).

Kiladze crater[2] (lat. 28.4° N, long. 212.9°, diameter ~50 km; Fig. 1) and surroundings northeast of Sputnik Planitia on Pluto display a prominent local concentration of $H_2O$ ice (Cook et al. 2019; Emran et al. 2023). In the case of some other exposures of $H_2O$ ice on the dwarf planet, for example in the Virgil Fossae and Viking Terra regions (see Fig. 1), there is also a red-orange pigmentation of the region approximately coincident with the $H_2O$. However, the coloration at Kiladze is somewhat less vivid than in the two examples cited. The discovery of ammonia ($NH_3$) or another ammoniated compound in spectra of the pigmented $H_2O$ ice at Virgil Fossae and Viking Terra obtained with the LEISA mapping spectrometer[3] on the New Horizons spacecraft (Dalle Ore et al. 2019) led to the proposition that relatively recent cryovolcanic activity has occurred in these regions (Cruikshank et al. 2019a, 2020).

Both at Virgil Fossae and the Viking Terra complex, the presence of $NH_3$ or an unidentified ammoniated compound suggests a relatively young age for the emplacement of the $H_2O$ in which it is carried. The chemical identification of the ammoniated compound is ambiguous because the absorption band at 2.21 μm detected in the LEISA spectra (Dalle Ore et al. 2019) is common to $NH_3$ ice as well as some ammoniated minerals (Berg et al. 2016), salts (Fastelli et al. 2020), and ammonia hydrates (Schmitt et al. 1998; Cruikshank et al. 2005; Bertie & Shehata 1985). Ammonia ice is readily destroyed by radiation sources in Pluto's space environment, while the spectral signature near 2.21 μm in minerals and salts, and perhaps ammonia hydrates, is expected to be more durable (Loeffler et al. 2010a, b; Cruikshank et al. 2019a). Consequently, the presence of the absorption band is suggestive of the relative youth of the emplacements, while the actual age is very difficult to ascertain.

---

[2] Previously known informally as Pulfrich crater. The feature looks like an impact crater, though it lacks the typical morphology of an impact crater. Therefore, we choose not to refer to Kiladze as a crater, rather as a depression or a structure.

[3] Linear Etalon Imaging Spectral Array (LEISA) is a mapping spectrometer covering the wavelength range 1.25-2.5 μm with spectral resolution $\lambda/\Delta\lambda = 240$, and with spatial resolution dependent on the range to Pluto's surface.

The foregoing establishes the relevance of the possible detection of the ammonia spectral signature in the Kiladze region as it relates to the origin of the original material, its possible cryovolcanic emplacement, and the duration of exposure to the Pluto environment, that is, the age. In this paper, we analyze the LEISA spectral images of Kiladze area and surroundings in a search for the absorption band at ~2.21 μm using a machine learning technique that isolates the relevant spectral region spatially and enables a detailed examination of the absorption band. We then modeled the Kiladze spectrum using Shkuratov's (1999) radiative transfer algorithm to establish the contribution of an ammoniated component at that area of Pluto.

## 2. Kiladze area and surroundings

The Kiladze depression is located northeast of the large basin, Sputnik Planitia, on Pluto. Fig. 1 shows the major geologic provinces on the dwarf planet, including the Kiladze area in relation to other prominent topography in Pluto's northeast quadrant. The Kiladze depression, Supay Facula, and immediate surroundings are shown in Fig. 1b, c, and d, which represent the highest spatial resolution achieved by New Horizons in this region. Moreover, Fig. 1d was processed to show the distribution of the red-brown coloration in this region.

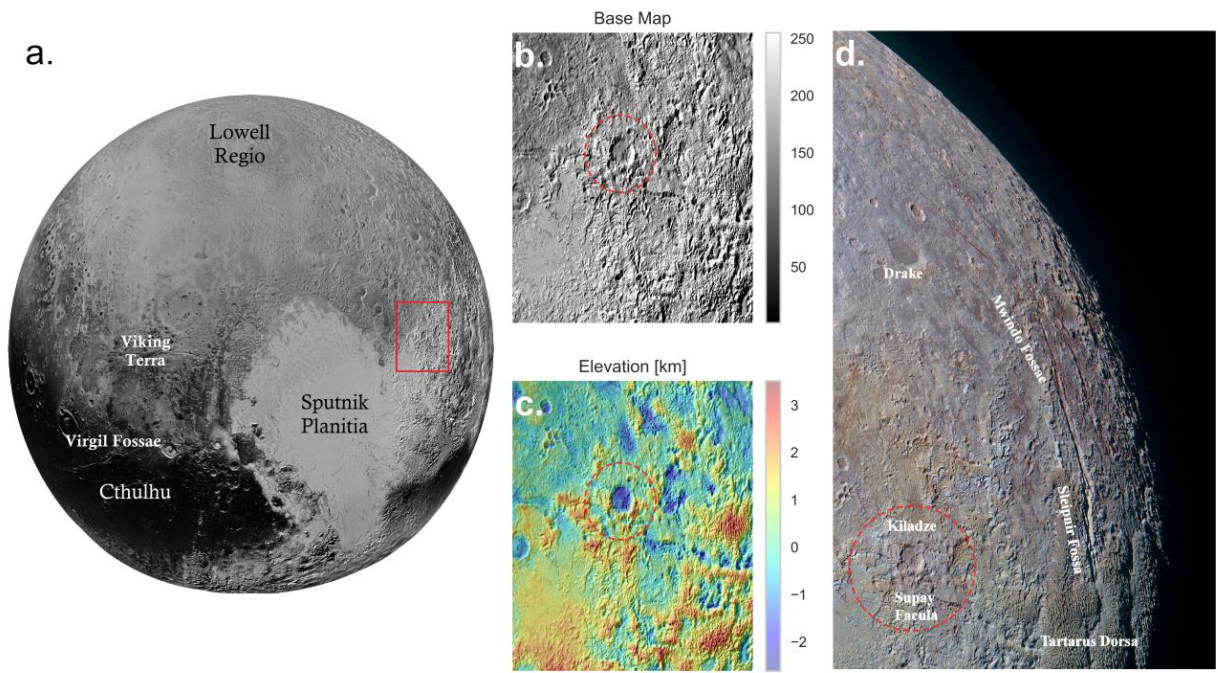

**Fig. 1.** The Kiladze depression and surrounding region on Pluto. Higher resolution LORRI basemap (a) shows major provinces including the areas where an ammoniated component was previously detected – Virgil Fossae and Viking Terra complex. The red rectangle indicates the location of the Kiladze area. The enlarged basemap of the study area (b), a color-coded topographic map of the Kiladze-Supay region (c; Schenk et al. 2018), and the distribution of red-brown coloration in Pluto's northeast quadrant using MVIC color mosaic (d). In each subplot, the red dotted circle indicates the location of the Kiladze area. Note that the highest elevations (c) in the lower part of this view correspond to the bladed terrain that is prominent further south in Tartarus Dorsa (Moore et al. 2018).

A large-scale compositional map is useful for putting Kiladze in context with the region. Using spectral images from the LEISA mapping spectrometer on New Horizons, maps of the principal components of Pluto's surface, specifically $H_2O$, methane ($CH_4$), and nitrogen ($N_2$) ices, plus the red-orange pigment, can be overlain on the highest resolution MVIC images. Here we present images of the Kiladze-Mwindo-Sleipnir region showing the distribution of $H_2O$ and the pigment because of their mutual association over much of the planet's surface.

Fig. 2a is a map of the distribution of $H_2O$ ice across the region of interest, based on the 2.0-μm absorption band depth. Orange represents the greatest band strength, hence the greater concentration of $H_2O$. Green represents a slightly weaker band, while blue maps the weakest band strength. Water ice is most abundant in Kiladze area and the nearby surroundings to the south and east, as well as in a few isolated patches. A lesser band strength is found in the trenches of Mwindo Fossae and the portion of Sleipnir Fossa that extends into Mwindo. Other exposures of $H_2O$ occur in several crater floors and some of the adjacent terrain throughout the area. The higher elevations of the bladed terrain in Tartarus Dorsa are essentially devoid of $H_2O$ ice. In the Mwindo Fossae complex, the LEISA maps have insufficient spatial resolution for a clear identification or rejection of the band, and the conclusions we reach about this region are based on inference from Kiladze and the exposures at Virgil Fossae and Viking Terra.

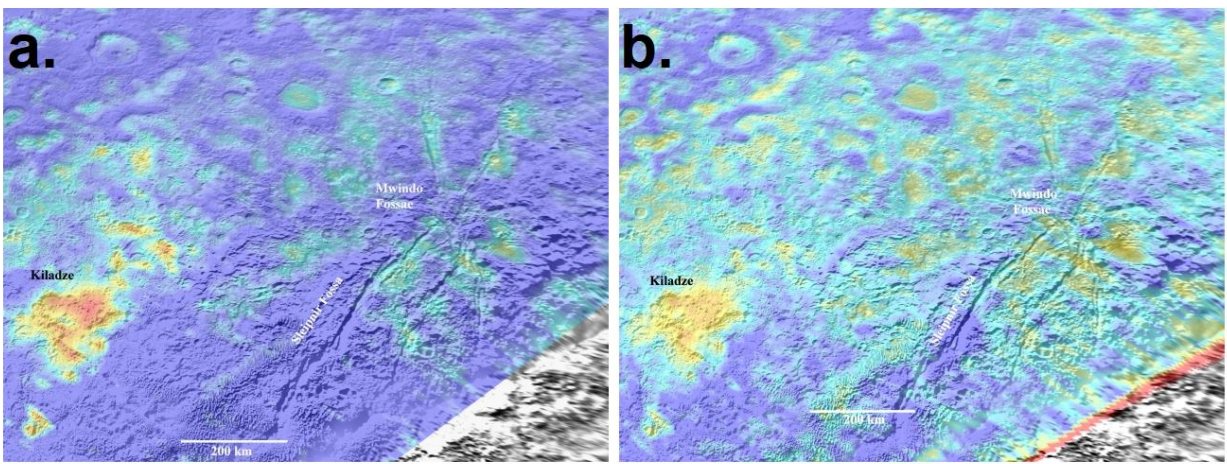

**Fig. 2:** Distribution of $H_2O$ ice (a) and red-orange color (b) in the Kiladze-Mwindo-Sleipnir region. Orange and yellow represent the strongest concentration, while purple represents the least. The coloration is widespread over this region, with the most intense in and around Kiladze. Some coloration occurs in the trenches of Mwindo Fossae, consistent with the color-enhanced image in Fig. 1d.

The corresponding map of the distribution of the red pigment is shown in Fig. 2b. The strength of the coloration is gauged by the degree of separation of its spectral slope from the signatures of both $H_2O$ and $CH_4$ containing ices, as described by Schmitt et al. (2017). The coloration is seen to be strongest in Kiladze and nearby terrains, and in a patchy distribution in the Mwindo Fossae troughs and surroundings. The various red-orange and dark brown colors of Pluto may have different origins. Grundy et al. (2018) propose that precipitation of atmospheric haze particles over the planet's age gives color to the surface, accounting for the color variations by chemical reactions with indigenous ices, which are themselves variable with seasonal cycles of condensation and sublimation and secular changes. Variable particle precipitation across different regions of the surface may also yield color differences.

The coloration in this region of Pluto's surface likely arises from energetic processing surface ices, notably $CH_4$ and other possible hydrocarbons, by the solar wind, ultraviolet light, and cosmic rays. These energy sources, insofar as they reach the solid surface, can produce complex organic molecular material that is characteristically yellow, red, or brown in color (see Cruikshank et al. 2021, Fig. 16). The solar wind is mostly deflected and does not reach Pluto's surface and may not be a significant energy source effecting chemical changes in the ices. The degree to which Pluto's atmosphere is transparent to ultraviolet light from the Sun and interplanetary space is expected to be variable over seasonal timescales as the abundance of methane changes (Bertrand et al. 2019). The thin atmosphere is entirely transparent to cosmic rays. A model of cosmic rays incident on Pluto's $H_2O$ ice-rich surface predicts that the energy deposition in the uppermost meter is ~$10^7$ eV/g/s (Cruikshank et al. 2019), sufficient to induce chemical changes in trace constituents of the ice such as $CH_4$, $NH_3$, CO, and $CH_3OH$. Irradiation of a mix of ices ($N_2$, $CH_4$, CO) by 1.2 keV electrons in a laboratory setting produced a strongly colored refractory residue rich in organic molecules that can be presumed to occur in some regions of Pluto's surface (Materese et al. 2015; Cruikshank et al. 2016).

A third source of highly localized red-orange pigment in Pluto's Virgil Fossae and Viking Terra may be organic-rich fluids debouched from subsurface reservoirs along faults that define graben structures that have resulted from regional stresses (Cruikshank et al. 2020). Fluids carried to the surface by cryovolcanic processes may also be colored by minerals resulting from the interaction of liquids with the rocky material that comprises much of Pluto's mass. There is no direct spectroscopic evidence for either the organic or mineral composition of the red-orange materials found at Virgil Fossae or Viking Terra, but the detection of an ammoniated component of the $H_2O$-rich ices there (Dalle Ore et al. 2019; Cruikshank et al. 2021) favors the presence of an organic component.

## 3. Ammonia analysis

### 3.1. Principal component reduced Gaussian mixture model (PC-GMM)

We isolate the spectral signature of the Kiladze area from LEISA image data to decipher its surface composition using the principal component reduced Gaussian mixture model (PC-GMM).

The same algorithm/ approach has successfully been used to map the distribution of global surface composition on Pluto using LEISA datasets (Emran et al. 2023). Inherently, PC-GMM is an unsupervised machine learning technique that first utilizes the principal component analysis (PCA) to reduce data dimension, then an unsupervised Gaussian mixture model (GMM) to group pixels by their signatures. We refer readers to the original paper of Emran et al. (2023) for a detailed description of the analytical technique of PC-GMM. Here we apply the PC-GMM to the Kiladze region, including the adjacent Supay Facula, as shown in Fig. 1b.

In this instance, we use the derived products of the LEISA dataset (Stern 2018). The image scene can be accessed from the PDS Small Body Node[4]. The image was processed and calibrated with the mission data pipeline processing routine including the bad pixel masking, background noise cleaning, flat fielding, and conversion of DN to radiance factor (*I/F*) modules. Details of the image processing routine are found in Schmitt et al. (2017) and Protopapa et al. (2017). The calibrated image was then projected to an orthographic viewing geometry using the United States Geological Survey's (USGS) Integrated Software for Imagers and Spectrometers (ISISv3) software package. The details of the image scene used are given in Table 1.

**Table 1.** Details of the LEISA scene used in this study.

| MET | Scan name | UT date and time | Range (km) | Sub S/C Lon (°) | Sub S/C Lat (°) |
|---|---|---|---|---|---|
| 0299172014 | P_LEISA_Alice_2a | 2015-07-14 09:33:05 | 112742 | 158.62 | 38.52 |

*Note:* MET = Mission Elapsed Time; S/C = Spacecraft

Further processing of the LEISA data includes an incident angle calibration as implemented by Schmitt et al. (2017) and Emran et al. (2023). However, in this work, unlike Emran et al. (2023), instead of linear interpolation we used cubic interpolation to fill in the missing (no data) pixel values. We choose this approach since the image subset used in this study covers a smaller spatial area (compared to the entire global disk used in Emran et al. 2023) and in many aspects, a cubic interpolation renders a much smoother output compared to a linear one. The image was also calibrated with the LEISA radiometric scaling factor of 0.74±0.05 (Protopapa et al. 2020) and excluded all pixels that correspond to an incident angle of greater than 85° (Schmitt et al. 2017). For details of the complete image calibration treatment, we refer readers to Emran et al. (2023).

---

[4] https://pds-smallbodies.astro.umd.edu/index.shtml

Following the image calibration, we applied a principal component analysis using the scikit-learn (Pedregosa et al. 2011) python module. We find that the first four axes of the principal component (PC) encompass considerable surface information of the area – shown by the precise correspondence with the underlying morphology (and contrast) of the region (see Fig. A1). The pc-axis #5 onward shows "noise"[5] and thus we didn't include these higher pc-axes for further analysis. Moreover, the scree plot of the eigenvalues and the explained variance plot show that the first four pc-axes have substantial image information and cover more than 82% of the total variance (see Fig. A2). Further evidence comes from the nearly constant size of the eigenvalues after number #4 (right panel in Fig. A2). Thus, taking into consideration pc-axes, scree plot, and explained variance plot, we consider the first four pc-axes for the application of the GMM.

We choose a total of six clusters for the Kiladze area and surroundings based on Akaike information criterion (AIC) and Bayesian information criteria (BIC) values[6] and their gradient measurements. We refer the reader to Appendix (Section A3) for details of the approach used in choosing the "optimal" number of clusters in this instance. For the convenience of interpretation of the surface units distribution, we hereafter refer surface unit/cluster as C followed by the assigned surface unit number. The spatial distribution of the resulting surface units shows that the Kiladze depression belongs to cluster C6 with a few isolated patches north and south of the feature (Fig. 3). We also calculate the predicted likelihood of each unit at the LEISA pixel level based on estimated component density. The predicted probability scale for each pixel ranges from 0 to 1 – the higher values the larger the probability for the corresponding surface unit at that pixel. The probability subplots for resulting clustering units also confirm that the likelihood of being cluster C6 is highest (around 1) in the Kiladze area and some isolated spots (see Fig. A4).

---

[5] We assign "noisy" data subjectively through visually analysis whether resultant pc-axes represent an appearance of the underlying geology (and contrast) seen in the basemap (Emran et al. 2023).

[6] For details of the AIC and BIC measures and rules of their interpretations, we refer the readers to the original paper by Emran et al. (2023).

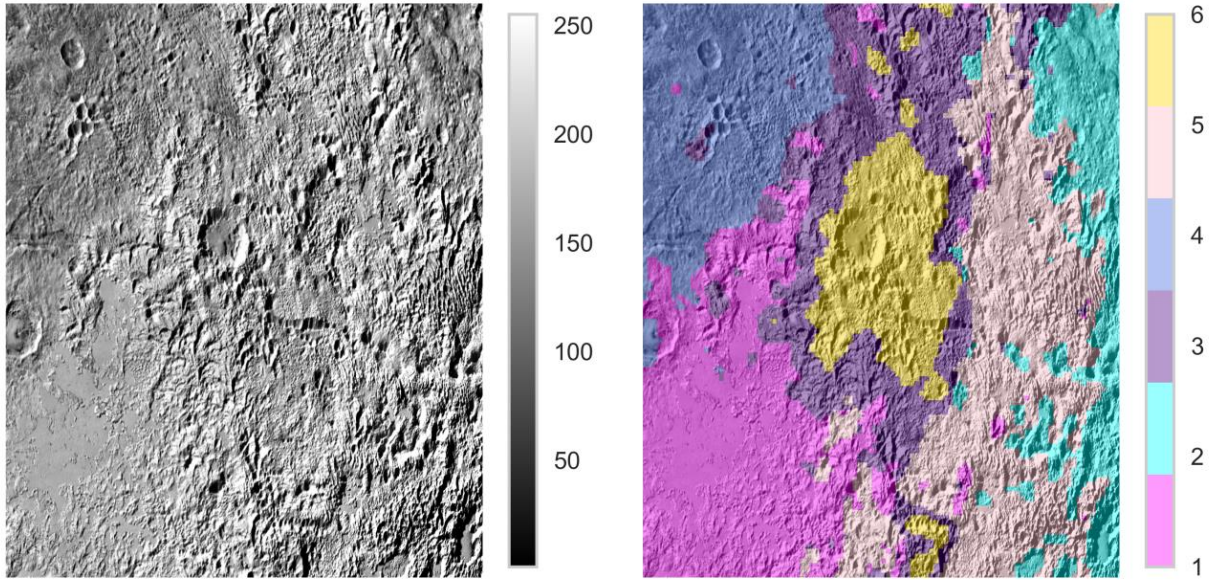

**Fig. 3:** Six surface units calculated for Kiladze depression and surrounding areas using PC-GMM. The right panel shows the spatial distribution of the surface units in the study area. The values in the legend indicate the assigned surface unit number. The geographical distribution of cluster C6 (yellow unit) is concentrated in the Kiladze area with a few isolated patches north and south of the Kiladze area. On the left panel, we added the higher resolution basemap of the same region for convenience when making a direct comparison with the area morphology.

The next step involved selecting the cluster(s) that showed the presence of $H_2O$ ice and discarding the remaining others. To this end, we retrieve the mean and 1-σ standard deviation of the I/F spectra from all the pixels that fall into each surface unit (Fig. 4). Similar to cluster number, the spectral subplot for the corresponding surface unit is labeled as C followed by the assigned unit number. Among the cluster averages, we note that spectra of C6 (Kiladze) show the typical spectral signature of $H_2O$ ice mixed with $CH_4$, i.e., a depression at ~1.5 µm and ~2.0 µm. In contrast, the nearby units, such as C1 or C5, exhibit rich in $CH_4$ as seen by having strong $CH_4$ absorption bands but lacking $H_2O$ bands. It is also noted that the spatial span of C3 follows C6 (see Fig. 3 and Fig. A4) with less intense $H_2O$ absorption features at ~1.5 and ~2.0 µm (Fig. 4). This disposition is suggestive of a gradient-like placement where the purer ice is located in the center i.e., cluster C6 and gradually fades outward. The spectra of C6 also show a broad absorption band around ~2.21 µm – which is an indication of the potential ammoniated materials. However, $CH_4$ ice also has a relatively weak and symmetric absorption band at ~ 2.20 µm. Consequently, a confirmed detection of ammoniated materials requires further analysis of the spectra. We selected the average spectra of all pixels that belong to C6 for additional analysis.

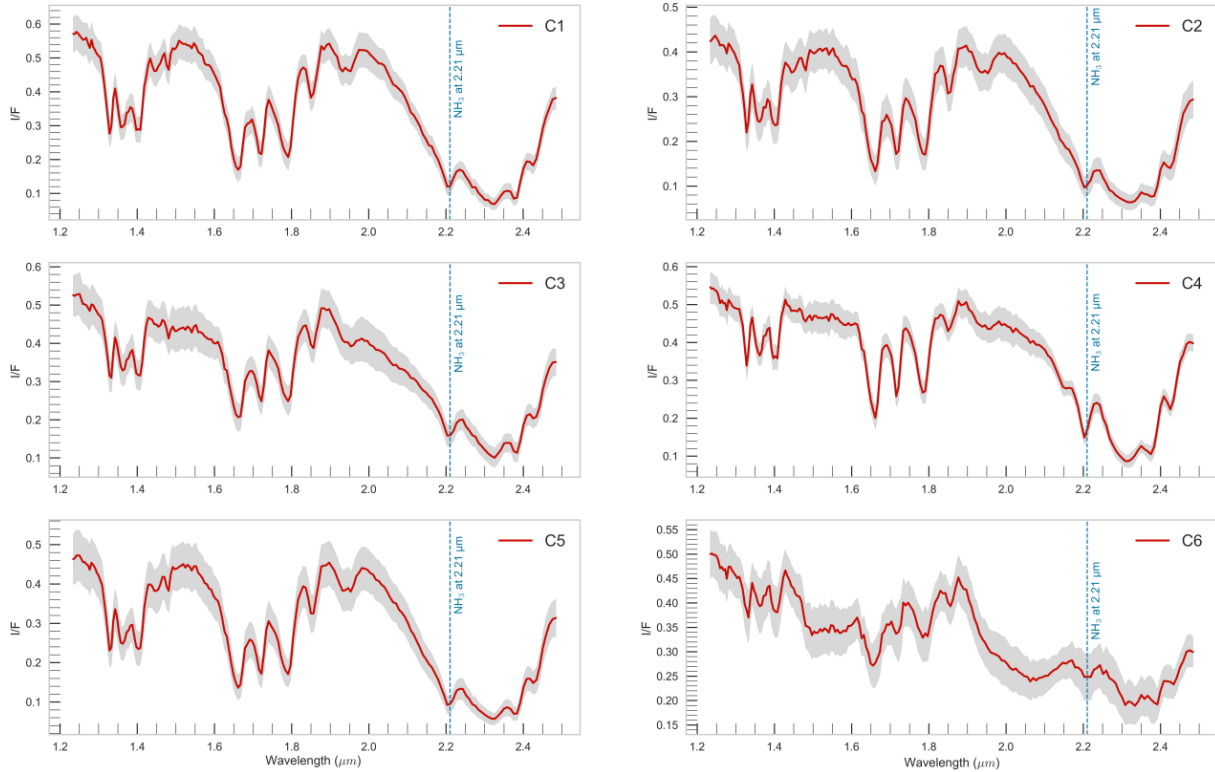

**Fig. 4:** The mean (red line) ± 1σ standard deviation (gray shade) I/F spectra of each surface unit. The standard deviation of the I/F spectra follows varying degrees of closeness to the mean spectra at different LEISA wavelengths. The spectra of the surface units in the subplots are labeled as C followed by the corresponding assigned surface unit number. C6 shows absorptions at ~1.5 and ~2.0 µm, corresponding to the $H_2O$ spectral signature, and a broad absorption band around ~ 2.21 µm – an indication of the potential ammoniated materials. This band is also seen in the extracted spectrum of the Kiladze region (identified therein as Pulfrich) shown in Fig. 4 of Schmitt et al. (2017). The vertical cyan dotted line in each subplot indicates the location of 2.21 µm.

As seen in Fig. 4, all cluster averages with the exception of C6 show a sharp absorption at a slightly shorter wavelength than the 2.21-µm vertical line. On the other hand, the C6 spectrum hosts a broad absorption band around 2.21 µm where the center of the absorption band is not readily identifiable from visual inspection. Therefore, we attempt at measuring the center and the depth of the broad absorption band around 2.21 µm for C6 spectra. We use *cana*[7], an open-access python module, to calculate the band depth of the Kiladze spectra around the 2.21-µm depression. We model a continuum between 2.15 – 2.24 µm at a small wavelength interval (a window of 0.001 µm) to ensure a finer resolution fit to the Kiladze spectra. In this instance, we utilize the continuum fit model with a minimum of 3-σ levels for absorption band detection. The model continuum fit indicates that the Kiladze spectrum has a band depth center at 2.21±0.0012[8] µm with a depth is 7.81±0.693 % (Fig. 5) – an indicator of potential (positive) ammonia deposition. However, this analysis also cannot definitely confirm whether the 2.21-µm absorption band here is indeed due to the contribution from ammoniated components, therefore, warranting further analysis.

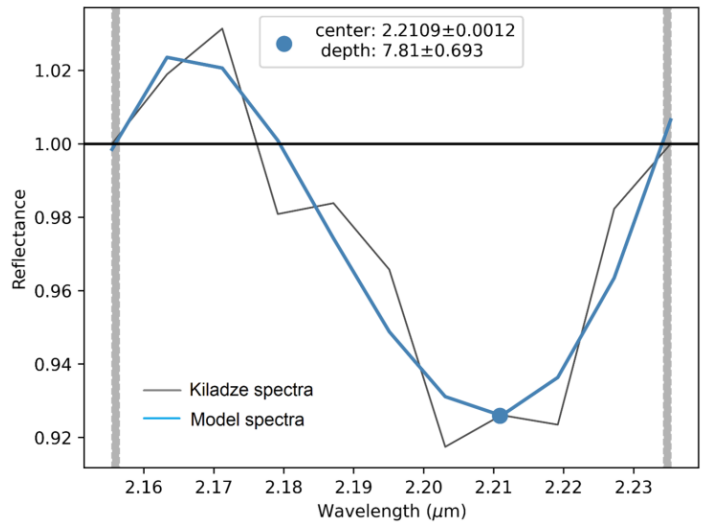

**Fig. 5:** Center of absorption bands between 2.15 - 2.24 µm for the Kiladze spectra (C6). The modeled spectra indicate that the center of the absorption band is 2.21 µm with a very small error margin (0.0012). The band depth is measured in percentage shows 7.81±0.693 %.

Since the Kiladze area shows red material contamination (Fig. 1d), we investigated the potential influence of the red coloration on the spectral absorption at 2.21 µm. We use a ratio analysis similar to Dalle Ore et al. (2019) where a ratio of cluster average to a standard spectrum was applied to identify the presence of ammoniated products in the Virgil Fossae region in Cthulhu. In that case, the cluster average that showed the least amount of ice was adopted as the standard against which all other cluster averages were measured, including that corresponding to Virgil Fossae (VF) itself. Here we adopted the same cluster average belonging to the region in Cthulhu as standard and took the ratio of that spectrum to those of C6. The resulting ratios are shown in Fig. 6a compared to the results from the Dalle Ore et al (2019) work for Virgil Fossae and the average spectrum of Pluto's small satellite Nix (Cook et al. 2018). When comparing the Kiladze ratio spectra belonging to this work to those of Nix and Virgil Fossae a few discrepancies become evident.

---

[7] De Pra, M., Carvano, J., Morate, D., Licandro, J. Pinilla-Alonso, N. (2018). CANA: An open-source python tool to study hydration in the Solar System (available at https://github.com/cana-asteroids/cana).

[8] 1-σ error bar (standard deviation)

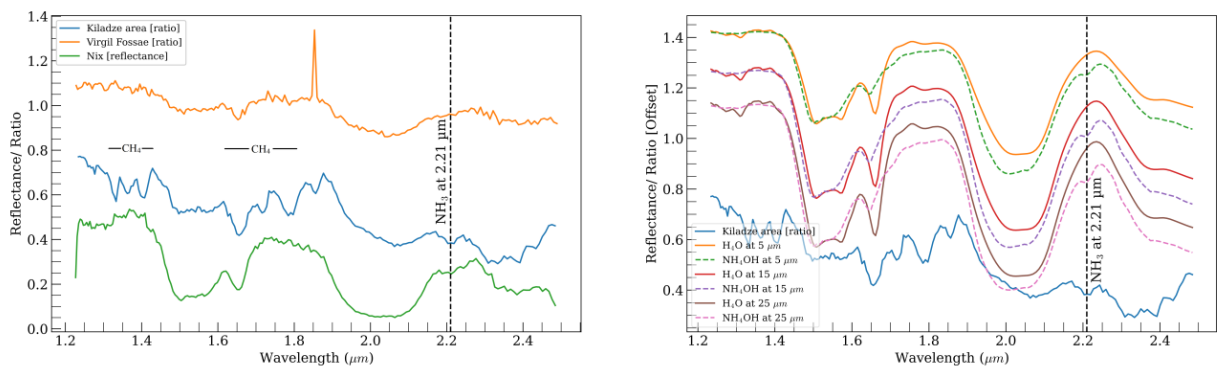

**Fig. 6:** (a) Comparison of the ratio spectra of the Kiladze area, ratio spectra of Virgil Fossae from Dalle Ore et al. (2019), and the spectrum of Pluto's small satellite Nix (Cook et al. 2018). The Kiladze ratio spectrum shows traces of $CH_4$ prominently lingering at ~1.3-1.4 µm bands. (b) Visual comparison of the Kiladze ratio spectra with 5, 15, and 25 µm grains of $H_2O$ and $NH_4OH$ spectra using the Shkuratov model (1999). The Kiladze ratio spectra exhibit similarity with the $NH_4OH$ spectra, compared to pure $H_2O$ spectra at ~ 2.21 µm.

Note that not only does the Kiladze spectrum (C6 in Fig. 4) have weaker $CH_4$ bands at 1.3-1.4 µm and 1.65-1.8 µm, but the ratio spectra shown in Fig. 6a indicate the presence of these absorption bands. However, traces of $CH_4$ stand out prominently in the ratio at ~1.3-1.4 µm. There is also a stronger than expected 1.65-µm band, which can be attributed to both $H_2O$ and $CH_4$, and a small but evident band at 2.2 µm that could be attributed to both $CH_4$ and ammoniated products. Finally, there is an absorption at ~2.3-2.4 µm that is of unknown origin.

When compared to the ratio spectrum of Virgil Fossae (Dale Ore et al. 2019) in Fig. 6a, the story is different. Some of the features described above are also present (though with less intensity) in that spectrum – the bands at ~1.3-1.4 µm, the stronger than expected 1.65-µm band, and the weak band at 2.2 µm. A spurious spike occurs ~1.85 µm in the spectrum of Virgil Fossae. The striking discrepancy between the two spectra (Kiladze vs Virgil Fossae) is in the depth of the band at 2.21 µm – Kiladze indicates a substantially stronger absorption at this wavelength. Another noticeable difference between these two ratio spectra is in the continuum level at ~2.3 µm. In the Virgil Fossae ratio spectrum, the height of the continuum is slightly lower than that at ~1.85 µm and is slightly negative, while the continuum slope is much more negative in the case of the Kiladze spectrum. In Dalle Ore et al. (2019) the depression in the whole spectral region between ~2.2 and 2.3 µm was interpreted as an indication of the presence of ammoniated products (salts and ions?), and we can reasonably assume the same here.

To better understand the spectral behavior in this region of Pluto's surface, we compare the Kiladze ratio spectra with signatures of pure $H_2O$ ice at 40K with that of $NH_4OH$ diluted in $H_2O$ (Fig. 6b). For this comparison, we computed model spectra using Shkuratov et al. (1999) approach and calculated for two different ices: one with pure $H_2O$ ice at 40 K (solid trace) and the second using $NH_4OH$ that had been diluted in $H_2O$ in the laboratory. To this end, we adopted the optical constants of $H_2O$ ice from Mastrapa et al. (2008, 2009) and $NH_4OH$ ice from Brown et al. (1988). In this instance, we model spectra at the three different grain sizes – 5, 15, and 25 µm for $H_2O$ and $NH_4OH$. A porosity parameter value of 0.7 was used for the Shkuratov model (1999) in this instance. We choose this value since a porosity of >0.4 has been hypothesized on the two Galilean icy moons Ganymede and Calisto (Black et al. 2001), while Europa's surface porosity may be up to 0.9 (Hendrix et al. 2005).

The resulting spectra (Fig. 6b) show the change of absorption bands due to grain size differences. Our main objective in modeling the effect of grain size was to observe the behavior of the height of the continuum on the two shoulders of the 2.0-µm band. Figure (6b) shows that the levels of the left and right shoulders of the 2.0-µm band are unchanged by variations in grain size in the range of moderate grain sizes that are commonly present on the surface of Pluto. However, there is an apparent change in continuum height on the right side of the 2.0-µm band for those models containing $NH_4OH$ diluted in $H_2O$. This supports our original claim of the presence of ammoniated products on the surface of Kiladze area. Even in this case, very little change is observed among the different grain sizes.

We visually compare the shape of the modeled spectra at 2.21 µm with the Kiladze ratio spectra. The results show that the Kiladze ratio spectra exhibit similarity with the $NH_4OH$ spectra compared to pure $H_2O$ spectra at ~2.21 µm band. Though this visual comparison gives a positive impression, this evidence cannot concretely affirm the presence of ammoniated materials at the Kiladze, until other parameters are known, for instance, component abundance. Thus, in the following section, we fit a linear model for spectra estimated using the Shkuratov (1999) approach to the Kiladze spectra to estimate the relative abundance and grain sizes of each constituent element on that part of the surface.

**3.2. Model**

To explore if there is indeed an influence of ammoniated materials in this region of Pluto's surface and estimate its fractional contribution, we model the spectra using the widely adopted radiative transfer model of Shkuratov et al. (1999). We refer readers to the original publication associated with the model for the algorithm's details. To that end, we use the optical constants of pure crystalline $H_2O$, $CH_4$, and $NH_4OH$ ices and the average reflectance of Cthulhu macula area. We employ the optical constants of $CH_4$ ice since we observe residual traces of $CH_4$ absorption bands in both Kiladze spectra and its ratio. $CH_4$ optical constants were adopted from Grundy et al. (2002). Details of the optical constants used are described in Table 2. For the average reflectance spectra of Cthulhu, we used the average spectra extracted from the entire Cthulhu region by Emran et al. (2023). We chose the average spectrum of the entire Cthulhu region because the coloration in the Kiladze area is not well identified from the compositional viewpoint with that of any specific part of Cthulhu. Moreover, using an average spectrum of Cthulhu minimizes unwanted errors since the spectra show a near absence of the absorption features of the known ices on Pluto (see Fig. 8 of Emran et al. 2023 for Cthulhu spectra).

**Table 2**. Optical constant used in this study.

| Materials | References | Wavelength range (µm) used | Notes |
|---|---|---|---|
| $CH_4$ | Grundy et al. (2002) [a] | 1.2 – 2.5 | **$CH_4$**:$N_2$ [b] |
| $H_2O$ | Mastrapa et al. (2008, 2009) | 1.2 - 2.5 | Crystalline phase |
| $NH_4OH$ | Brown et al. (1988) | 1.2 – 2.5 | $NH_4OH$ was diluted in $H_2O$ |

[a] Optical constants are available at https://www.sshade.eu/

[b] Pure $CH_4$ ice optical constant at 39K used in this research can also be used as a proxy of $CH_4$ diluted in $N_2$ ice (**$CH_4$**:$N_2$; Protopapa et al., 2017; Emran and Chevrier 2022).

Using the Shkuratov model (1999), we calculate the reflectance spectra of component endmembers from the optical constants. For the red-materials endmember (Cthulhu spectra), we estimate the reflectance from the I/F spectra by dividing it with π (pi). As already mentioned, we adopt a value of 0.7 for the porosity parameter.

We fit a model using a (nonnegative) multiple linear regression approach — areal or geographical mixture — to estimate the fractional contribution of each component. The concept of areal mixing is based on the idea that surface composition consists of a mixture of different components (compositional and/or microphysical characteristics) in such a way that each component stays isolated spatially from the others (Protopapa et al. 2017). Thus, the multiple linear regression model used here assumes that the Kiladze spectra are contributed by the summed-up reflectance of individual component spectra weighted by their fraction areal contribution. We adopt this strategy since it is the simplest method and a similar strategy has successfully been applied to modeling the Pluto surface composition (Protopapa et al. 2017) and Martian dune composition (Emran et al. 2021). With the areal mixing model, the reflectance of a spectrum can be written as:

$$r_f = \sum_{i=1}^{n} f_i r_i; 0 \leq f_i \leq 1 \qquad (1)$$

where $r_f$ is the reflectance of the spectrum (Kiladze spectra in this instance), $f_i$ is the fraction contributed by the $i^{th}$ endmember, and $r_i$ is the reflectance of the $i^{th}$ endmember. Once the model is executed and best-fit parameters are estimated, the relative abundance ($A$) of each of the contributed components can be calculated as:

$$A_i = \frac{f_i}{\sum_{i=1}^{n} f_i} * 100 \qquad (2)$$

where $A_i$ is the relative abundance contributed by the $i^{th}$ end member (in percentage).

Note that the ultimate intended parameters to estimate using the multiple linear regression model are the relative abundance (%) and diameters (μm) of each constituent. We estimate the best fit of these parameters by adopting the least squares routine to fit the model to the Kiladze spectra using the Levenberg-Marquardt $\chi^2$ minimization algorithm (Levenberg 1944; Marquardt 1963). To validate if $NH_4OH$ has a contribution to the Kiladze spectra, we utilize the least square routine with and without the inclusion of the optical constants of $NH_4OH$. The best-fit parameters estimated from the linear model are listed in Table 3.

**Table 3:** Best-fit parameters from the multiple linear regression model.

| Parameters [a, b] | Without $NH_4OH$ [c] | With $NH_4OH$ [c] |
| --- | --- | --- |
| Diameter of $CH_4$ | 415.5 ± 67.18 μm | 367.39 ± 50.12 μm |
| Diameter of $H_2O$ | 30.95 ± 6.8 μm | 16.96 ± 92.07 μm |
| Diameter of $NH_4OH$ | -- | 20.04 ± 8.54 μm |
| Abundance of $CH_4$ | 9.74 ± 0.39% | 10.4 ± 0.37% |
| Abundance of $H_2O$ | 13.94 ± 0.67% | 1.24 ± 1.67% |
| Abundance of $NH_4OH$ | -- | 14.87 ± 1.61% |
| Abundance of red coloration | 76.32 ± 9.13% | 73.45 ± 9.62% |
| Chi-square ($\chi^2$) | 0.06 | 0.04 |

[a] Diameter in μm
[b] Relative abundance in percentage
[c] Mean best-fit ± 1-σ error

The result (Table 3) shows that the areal mixing model fits better when the model is fitted including the optical constants of $NH_4OH$ as it renders a lower chi-square $\chi^2$ value of 0.04, compared to 0.06, when the model fits without optical constants of $NH_4OH$. Consistent with red coloration as seen by the visual inspection of Fig. 1d, the model also confirms that the red coloration has the highest relative abundance (>70%) in the Kiladze spectra. This high contribution is independent of with or without including the $NH_4OH$ optical constants. The model also indicates that ~ 15% of the relative abundance (areal fraction) comes from $NH_4OH$ contribution. Note that a much lower amount of $H_2O$ needed when using ammonia is probably due to the fact that the $NH_4OH$ constants are for ammonia diluted in water and the total amount of $H_2O + NH_4OH$ is about constant in the two cases.

Table 3 indicates the difference between the mean diameter of $H_2O$ ice in both cases is not much. However, the standard deviation of the $H_2O$ diameter is better without ammonia while there is a clear improvement in the $CH_4$ fit with the ammonia. A relatively higher standard deviation of the $H_2O$ diameter with ammonia may be an indication that the diameter of $H_2O$ ice does not play much influence in the model fit (also the abundance is much lower). This is may also due to the influence of $NH_4OH$ which itself has a contribution from $H_2O$ ice – as mentioned earlier. However, amid the better model fit – a lower chi-square $\chi^2$ value over entire wavelengths — supports the detection of an ammoniated component at the Kiladze area.

We also plot the best-fit model to the data and resultant residuals at the LEISA wavelengths (Fig. 7). The plot also visually confirms that the areal mixing model fits better (lower residuals) when the model is fitted including optical constants of $NH_4OH$ (bottom panel of Fig. 6). Notably, the fitted model shows higher residuals — mean and $1\sigma$ standard deviation of residual of -0.00008 ± 0.0045 vs -0.00016±0.0056 — from Kiladze spectra at ~1.65 and ~2.21 μm when the model is implemented without the optical constants of $NH_4OH$. In contrast, at both 1.65 and 2.21 μm, the model fits better (lower residual) when the model is fitted to the Kiladze spectra with integrating optical constants of $NH_4OH$. Note that while the model shown here uses $NH_4OH$, a thorough analysis of the 2.21-μm absorption band on Charon (Cook et al. 2023) favors $NH_4Cl$ as the responsible material.

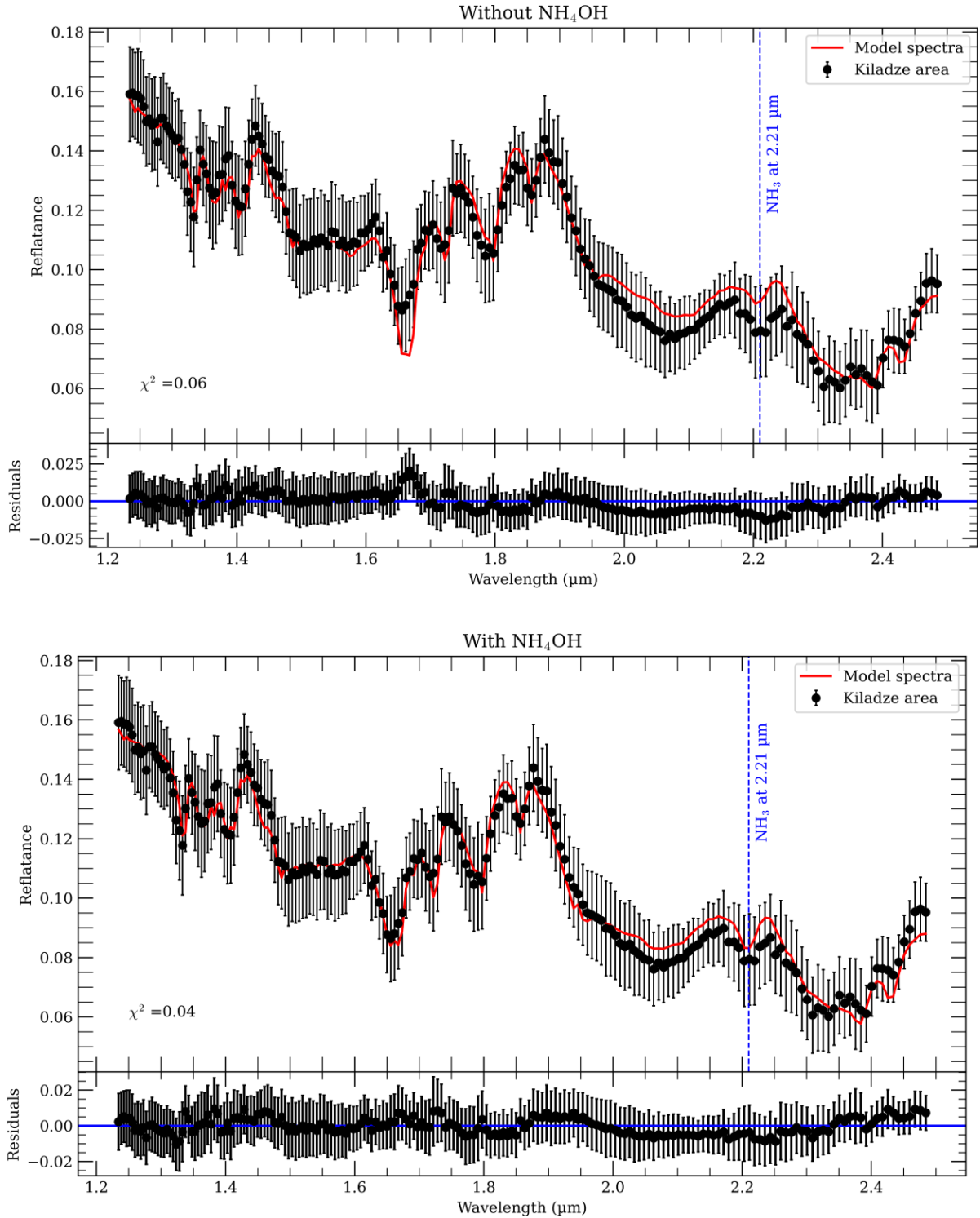

**Fig. 7:** The average and 1-σ standard error I/F spectra of the Kiladze area (black circles) and the best-fit model spectra (red line) using an areal mixing approach. The figure shows the multiple linear regression model fits better (lower $\chi^2$ values) when the model is implemented including optical constants of $NH_4OH$ (bottom panel) compared to without using $NH_4OH$ optical constants (bottom panel). The residual plot at the entire wavelengths is also included with each subplot.

## 4. Conclusions

We present results on the composition of a restricted region in the vicinity of the Kiladze depression on the eastern side of Sputnik Planitia based on a machine-learning technique focused on clustering large numbers of spatially resolved near-infrared (NIR) spectra. The technique was developed to extract endmembers in the spectral array, representing the purest possible diagnostic spectral band structure on different parts of Pluto's surface (Emran et al., 2023). We were successful at isolating the spectral signature of an ammoniated component of the basic $H_2O$ icy surface. The area appears less vividly colored, suggesting lower contamination from the reddish material that pervades other regions of Pluto and in particular Cthulhu. Previous works have identified the Kiladze region as rich in $H_2O$ ice (Cook et al. 2019; Emran et al. 2023) while being surrounded by areas heavily contaminated by $CH_4$ and its irradiation products (Protopapa et al. 2017; Schmitt et al. 2017).

Although the exact chemical identification of the surface component responsible for a band at ~2.2 μm remains undetermined, we demonstrate here that $H_2O$ at Kiladze carries a detectable signature of an ammoniated material. In our modeling, we use a mixture of $H_2O$ and $NH_3$ (ammonia hydroxide) as a placeholder for the precise chemical component. As noted above the options include ammoniated salts and minerals, in both of which the band is broad and unstructured, as well as ammonia hydrates. The wavelength of the band near 2.2 μm in the ammonia hydrates depends on the structure of the hydrate (e.g., Bertie and Shehata 1985). More generally, in mixtures of $H_2O$ and $NH_3$, the band position depends on the mixing ratio of the two components. Zheng et al. (2009) show that the 2.229-μm $NH_3$ band shifts to 2.208 μm as the percentage of ammonia decreases from 100% to 1%.

The spectral signature of the purest $H_2O$ ice in the Kiladze region shows similarities and differences with $H_2O$-rich spectra in the Cthulhu area previously analyzed by Dalle Ore et al. (2019). These areas were identified as contaminated by red material that is probably organic in nature, as well as ammoniated products. The main similarity between the spectra in Cthulhu with those in Kiladze is in the presence of a depression in the spectral region at 2.2 µm, attributed to the ammonia products. The most apparent difference is in the height of the continuum on the right shoulder of the 2.0-µm band, as discussed in a previous section. We have excluded the possibility that the difference could be due to a grain size effect, and we conclude that it is most likely related to the presence of ammoniated products in conjunction with the composition of the coloring pigment, which is thought to introduce an overall slope to the spectral continuum, as seen in some laboratory samples made by the irradiation of $CH_4$- bearing ices (Cruikshank et al. 2016, 2021).

Whereas at Virgil Fossae and Viking Terra, there are topographic structures indicative of fissures (graben) and evidence of flooded terrain that appear to support the cryovolcanism hypothesis, there are no corresponding geologic footprints (fissures/graben) in and around the Kiladze depression visible at the resolution of the available images that suggests fluid or cryoclastic emplacement. As noted, the large-scale crustal structure in the region east of Sputnik Planitia where Kiladze is located is significantly different from the structure west of Sputnik Planitia. Therein may be a clue to different causes for the compositional data presented in the two regions, while at the same time obscuring the sources and mechanisms by which emplacements of pigmented water ice bearing an ammoniated component have come about in two widely separated regions of Pluto's surface.

The assertion that the Kiladze structure represents cryovolcanism at this site on Pluto is not contingent on the detection of an ammoniated material in the $H_2O$ ice or the presence of a red pigment in the ice, although both of these conditions are met at the Viking Terra and Virgil Fossae locations. As noted above (Cruikshank et al. 2019a, b, 2021) the structural setting at Kiladze east of the Sputnik Planitia basin is different from that on the west side of the basin, and it is reasonable to assume that the chemistry and composition of the subsurface fluids feeding the putative cryovolcanic surface manifestations of cryovolcanism might be also different. As pointed out by Cruikshank et al. (2019, 2021), the chemistry of the cryomagma will be the result of the liquid $H_2O$, components of Pluto's rocky core, and other materials inherited from the feedstock in the outer Solar System where Pluto accreted and condensed. This may or may not have included ammonia or components that led to ammonia salts or other ammoniated compounds.

A major question about the overall color and composition of the Kiladze feature remains—how does the $H_2O$ icy surface remain exposed as a kind of bright oasis surrounded by the dominant $CH_4$ and $N_2$ ices that cover most of Pluto's landscape? These ices evaporate to the atmosphere and then recondense on seasonal and secular timescales, and when lying as a condensate on the surface are exposed to precipitating atmospheric aerosols composed of organic molecules that form as a consequence of ultraviolet sunlight and solar wind incident at high altitudes. Grundy et al. (2018) calculate that precipitating aerosol particles can accumulate to a depth of ~1 micrometer in one Pluto year (~3 earth years), and could form an opaque layer when it reaches few tens of micrometers in depth. In so far as some UV reaches the surface, seasonal deposits of condensed $CH_4$ ice are photolyzed to form other hydrocarbons, the more complex of which can become increasingly colored, or even opaque after some indeterminant period of time. In either case, it is to be expected that the pristine $H_2O$ uppermost surface of the Kiladze region would loose its high albedo, and perhaps the spectral signature of the ice in a relatively short time. In view of these factors, some mechanism must be at play to keep the surface highly reflective and the signature of $H_2O$ ice recognizable. If the region is volcanically active with frequent effusion of volatiles from the subsurface, this complicating factor might be resolved.

It is interesting to note that although the Kiladze depression resembles a heavily degraded impact crater, inspection of the highest resolution images indicates that the feature lacks the typical morphology of a crater (e.g., central peak). The floor of the depression is partly covered with a smooth unit, and has with multiple collapsed structures. Based on this evidence, we suggest that a water-rich cryolava carrying an ammoniated component may have come onto the surface at the Kiladze area via volcanic collapse, and the whole feature could be termed a volcanic caldera complex. This possibility, and others, will be examined in detail in further work in progress.

**Acknowledgments**

This work was supported by NASA's New Horizons mission. We express our gratitude to the mission operations personnel, the science team, and NASA management for their extraordinary contributions to the great success of New Horizons. C. Dalle Ore was supported in part by SETI Cooperative Agreement NNX13AJ87A. A. Emran was supported in part by the Univ. of Arkansas Center for Space and Planetary sciences. A. Emran also acknowledges the resources (e.g., computer) used from NASA Jet Propulsion Laboratory during the preparation of the manuscript. D. Cruikshank and J. Cook were supported in part by NASA's New Horizons mission of Pluto through the planetary surface composition team.

**Appendix:**

**Section A.1.** Principal component (PC-axes) of at Kiladze area and surrounding.

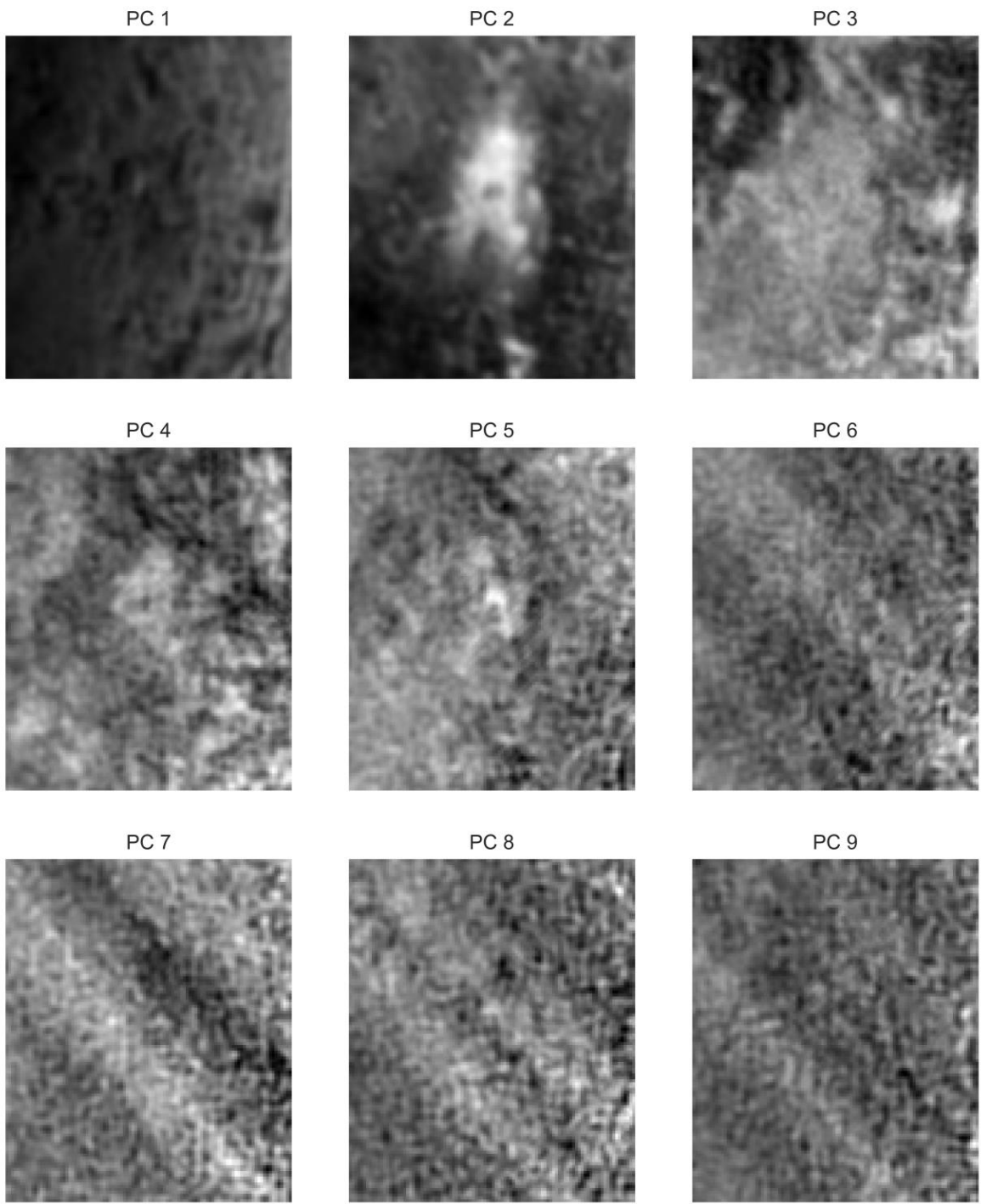

**Fig. A1:** The first nine principal components from the LEISA data cube at Kiladze area and surroundings. Up to PC4 show substantial surface information and covers 82% total variance. From PC5 and onward components show "noisy" data - do not clearly display the underlying geomorphology (and contrast) of Pluto in that area.

**Section A.2.** Cumulative explained variance and scree plot of pc-axes.

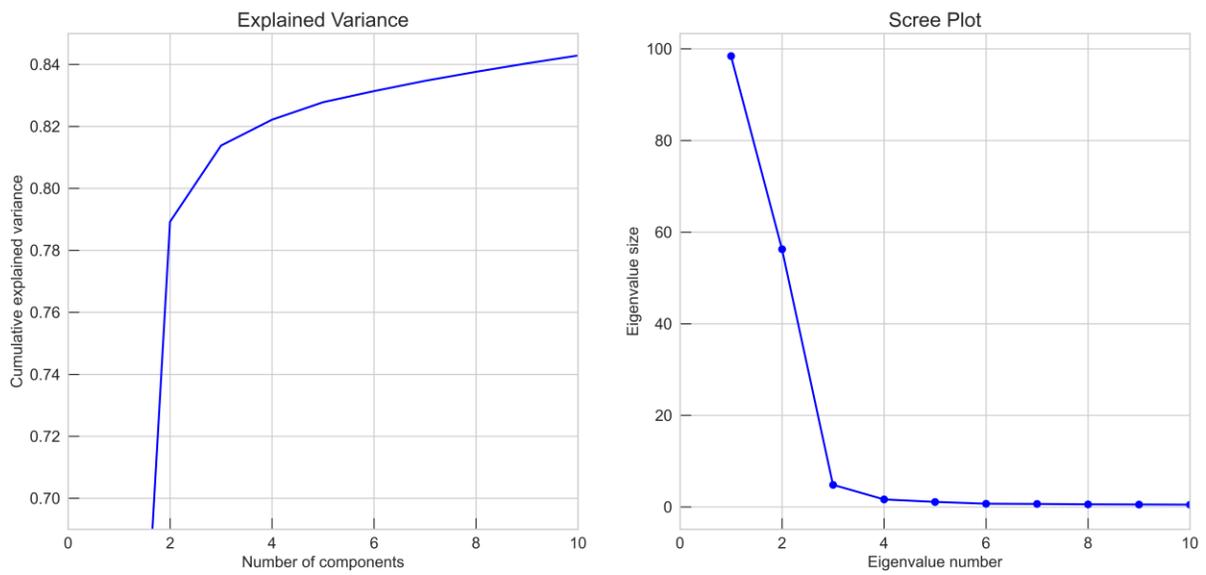

**Fig. A2:** The cumulative explained variance (left panel) and scree plot (right panel) for different pc-axes. The first four pc-axes comprise more than 82% total variance of the data.

**Section A.3.** Scaled AIC and BIC values and their gradients at the different number of clusters.

The optimal number of clusters ($N$) can be achieved from plots of AIC and BIC values at different numbers of clusters. Typically, the lower AIC indicates a better model fit (Liddle 2007) whereas the first local minimum of BIC value is considered the optimal number of clusters (Dasgupta and Raftery 1998; Fraley and Raftery 1998). Note that BIC values are also presented with an opposite sign by different studies (Fraley and Raftery 1998). Refer to Emran et al. (2023) for more details on the algorithm used for AIC and BIC and the selection criteria for the optimum number of clusters. Both AIC and BIC follow a similar trend as their values decrease with the number of clusters. Though BIC does not show any distinguishable local minimum, the gradients of both AIC and BIC show that when $N = 6$, there is a "local peak" of gradient values. Moreover, the gradients have abrupt change from $N = 6$; an opposite slope is seen and the gradient declines gradually up to $N = 7$. We postulate that no substantial information will be gained if the number of clusters is increased from 6. Thus, in this instance, we choose a total of 6 clusters for the Kiladze area and surrounding region on Pluto.

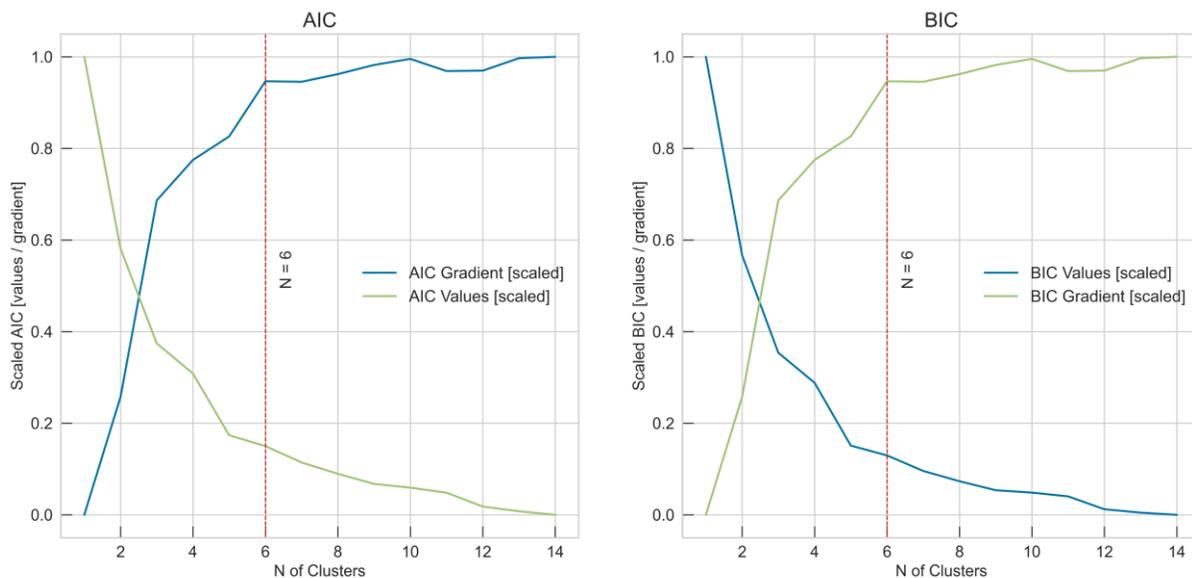

**Fig. A3:** Scaled AIC and BIC values and their gradients at the different number of clusters.

**Section A.4.** Probability plot of each surface unit at pixel label using PC-GMM at Kiladze area.

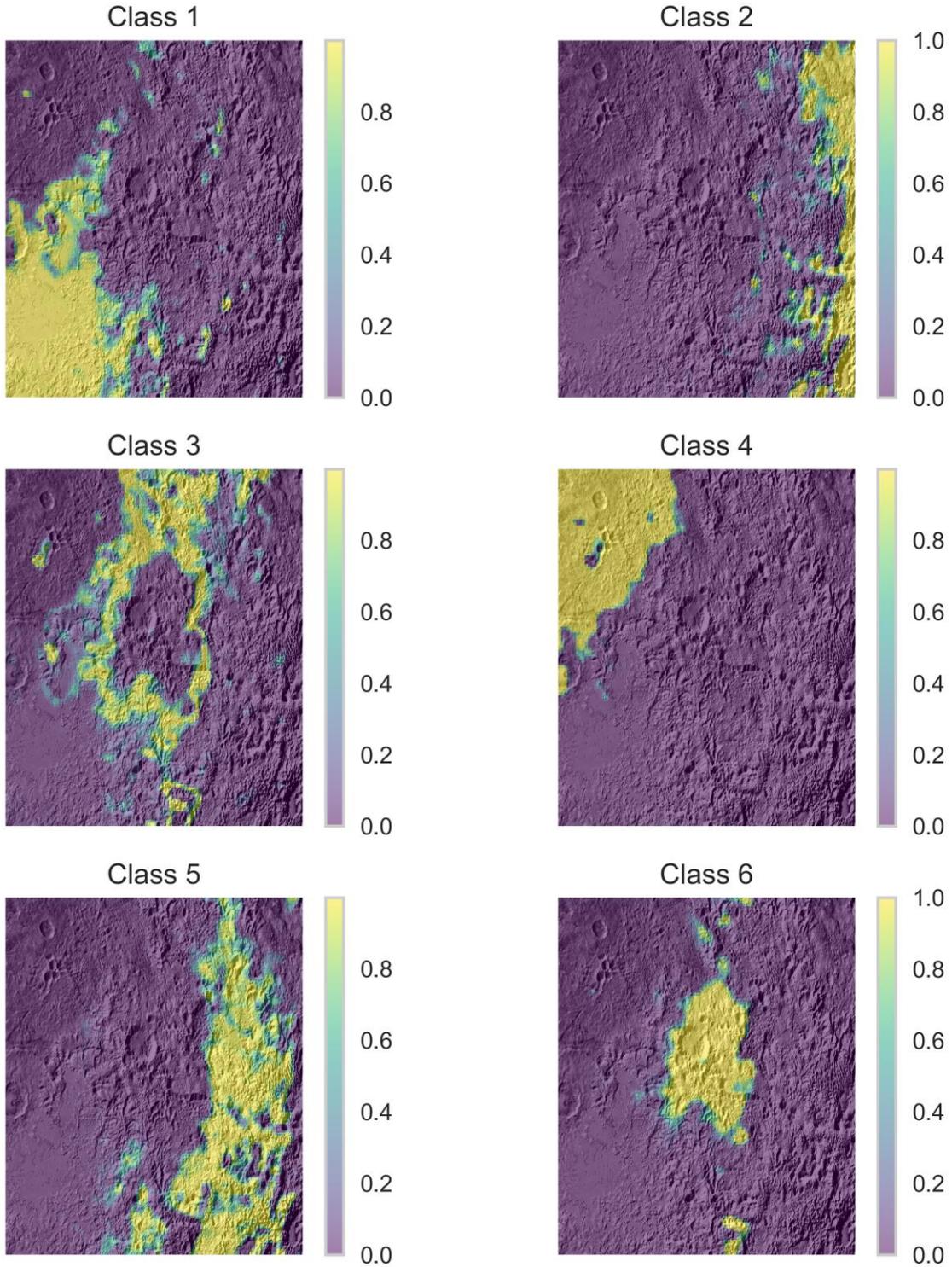

**Fig. A4:** Probability plot of each surface unit at LEISA image pixel level applied to the Kiladze area and surroundings. The probability scale ranges from 0 to 1 – higher values indicate a higher probability of the corresponding unit.